\documentclass[conference]{IEEEtran}
\IEEEoverridecommandlockouts
\usepackage{cite}
\usepackage{amsmath,amssymb,amsfonts}
\usepackage{algorithmic}
\usepackage{graphicx}
\usepackage{textcomp}
\usepackage{subcaption}
\usepackage{xcolor}
\usepackage{booktabs}
\usepackage{placeins}

\usepackage{tabularx}
\usepackage{array}
\usepackage{multirow}
\usepackage[hidelinks]{hyperref}

\def\BibTeX{{\rm B\kern-.05em{\sc i\kern-.025em b}\kern-.08em
    T\kern-.1667em\lower.7ex\hbox{E}\kern-.125emX}}
\begin{document}

\title{ERP Event Logs: A Canonical Form for KPI Time-Series
Extraction and Characterization}

\author{
\IEEEauthorblockN{
Sherri Hadian\IEEEauthorrefmark{1},
Adrian Rebmann\IEEEauthorrefmark{1},
Atacan Korkmaz\IEEEauthorrefmark{1},
Gregor Berg\IEEEauthorrefmark{1},
Ulf Brackmann\IEEEauthorrefmark{1}
}
\IEEEauthorblockA{
\IEEEauthorrefmark{1}\textit{SAP SE}, Germany \quad
\\
\{firstname.lastname\}@sap.com \quad
}
}

\maketitle\footnote{© 2026 IEEE. Personal use of this material is permitted. Permission from IEEE must be obtained for all other uses, in any current or future media, including reprinting/republishing this material for advertising or promotional purposes, collecting new collected works for resale or redistribution to servers or lists, or reuse of any copyrighted component of this work in other works.}
\begin{abstract}
Enterprise resource planning (ERP) systems record operations as transactions
spread across hundreds of normalized relational tables. Extracting an
operational key performance indicator (KPI) from this schema requires a
separate, bespoke join for almost every metric. We show that an event log, a
canonical (case, activity, timestamp) representation derived from the same
tables, collapses this heterogeneity into one flat structure, from which KPI families reduce to reusable operations after a one-time mapping. Using standardized event logs built
from hundreds of SAP S/4HANA customers under a shared activity ontology, we
define three families of process-derived KPIs (volumes, durations, and rates)
and extract them uniformly across customer systems for two common end-to-end
organizational processes---order-to-cash and procure-to-pay. We normalize
rates and screen series for minimum coverage to support comparison across
customers of very different sizes. We then characterize heterogeneity across
industries and tenants using descriptors derived directly from the canonical
form, before analyzing the resulting KPI time series. The same process executes
heterogeneously across customers and industries, yet this shared representation enables
direct comparison of sales- and procurement-side signals, including lead times
and order arrivals. Using this shared representation, we characterize which KPI
families are forecastable and find a recurring ordering from volumes through durations
to rates across both processes and a highly heterogeneous
customer base; a pretrained time-series foundation model (Chronos) is
competitive with classical baselines such as ARIMA and ETS on rate and duration
series, while classical models retain an edge on volumes.
\end{abstract}
\section{Introduction}
Operational data in ERP systems is transactional and highly normalized: a
single business document, such as a sales order or a purchase order, leaves traces
across many tables, each holding only slices of its lifecycle (e.g., creation,
delivery, invoicing, clearing). Computing an operational KPI over this schema means locating
the right timestamp columns in the right tables and writing a join tailored to
that metric. A different KPI needs a different join. There is no reuse: every
indicator is a bespoke query, and comparing indicators across customers, or
across two different processes, multiplies that effort. In practice, this
makes KPI extraction costly to engineer, fragile under schema
evolution, and difficult to standardize across customers and
processes.

\begin{figure*}[t]
\centering
\includegraphics[width=\textwidth]{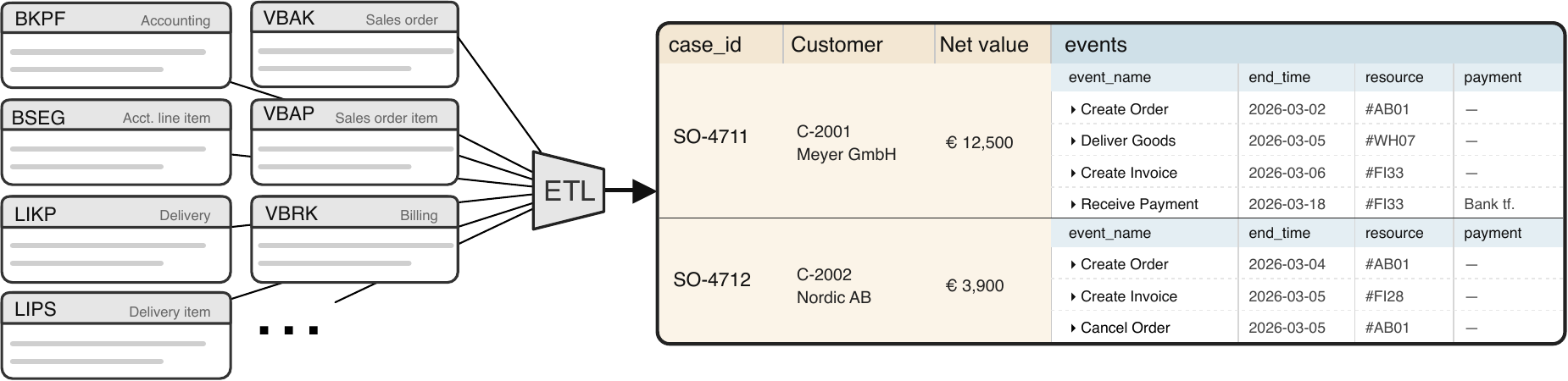}
\caption{From normalized ERP tables to the canonical form. A single business
document (a sales order) is spread across many tables. Headers and line items
for orders (VBAK/VBAP), deliveries (LIKP/LIPS), billing (VBRK), and accounting
(BKPF/BSEG). ETL reshapes them into an event log: each case carries its
case-level attributes and an ordered sequence of events that carry their own attributes. Every KPI is then a uniform operation over this one structure.}
\label{fig:canonical}
\end{figure*}

We observe that a single intermediate representation removes much of this cost. An
\emph{event log} records each document that marks the start of an end-to-end process execution as a case, and each processing
step as an (activity, timestamp) pair belonging to that case. Representing process executions as event logs is standard in process mining, which analyzes processes through case-linked activity and timestamp records~\cite{van2011process, van2022process, berti2021sapextraction}. Every ERP process, once mapped, reduces to this same (case, activity, timestamp) shape
(Fig.~\ref{fig:canonical}). 

Activity labels come from a fixed mapping that resolves each source record to
a canonical activity from a shared vocabulary, so the same business step is
named identically across customers. The source tables also differ in what they expose over time. Some carry a native
timestamp on each record---an order's creation date---from
which a series reads directly. Others store only current state and overwrite it,
so the timing of a rejection or price change survives only in the change history,
not in the state table itself. The event log places both kinds under one
time-resolved representation: series that already existed are extracted
uniformly, and those that the state tables had overwritten are recovered from the
change history. The event log carries no information that the source tables do not: it is derived from them,
and relational algebra could in principle recover any KPI from the normalized
schema. 
The gain is therefore not the expressive
power, which relational algebra already has, but \emph{extraction leverage}: 
 we pay this extraction cost once,
 and every KPI afterward becomes
an operation over one common structure rather than a new
schema-specific join. This yields reusable, comparable, and model-ready KPI time
series: the ontology aligns semantics, the common schema
reuses KPI operators, and the one-time mapping centralizes
source-specific joins.

Concretely, the process-derived KPIs we study fall into three
families, each corresponding to a common operational
question: \emph{how much}
work flows through a business process, \emph{how long} that work takes, and \emph{how often} it
deviates. A \emph{volume} is a count of an
activity per period, capturing demand and throughput (order arrivals,
completions); a \emph{duration} is the timestamp difference between two
activities in a case, capturing the
process lead times; and a \emph{rate} is a count of cases experiencing an event in a period, normalized by the active-case population, capturing the incidence of exceptions such as cancellations or price changes that signal process health.


To study the potential of this representation for KPI extraction and characterization, we build these logs from hundreds of SAP S/4HANA customers, repurposing the collection and mapping logic of \emph{SAP Signavio Process Intelligence}\footnote{SAP Signavio Process Intelligence is the process mining and analysis product from SAP SE, which we built on.} as the front end of a custom extraction pipeline. Its shared activity ontology supplies the cross-customer semantic
comparability, by applying a
fixed rule library over standardized source fields and coded
values to assign canonical activity labels.

We build event logs for two end-to-end processes---order-to-cash on the sales side and procure-to-pay on the procurement side---and use the shared representation first to characterize heterogeneity in the customer base through simple process descriptors, and then to extract comparable KPI time series for cross-process analysis and forecasting across tenants, illustrating reuse rather than directly measuring extraction cost.

Beyond the established event-log representation, our contributions are:
\begin{itemize}
\item We frame the event log as a canonical form over relational ERP
data, and argue its advantage is extraction leverage rather than added
expressiveness, illustrated by two worked examples
(Table~\ref{tab:leverage}).

\item We characterize process heterogeneity across hundreds of customers and industries using simple descriptors derived directly from the canonical form.

\item We define three families of process-derived KPIs---volumes, durations,
and active-case-normalized rates---as uniform operations over the log
(Table~\ref{tab:families}), comparable across tenants and across
order-to-cash and procure-to-pay.

\item We show the canonical form enables direct cross-process comparison of
sales- and procurement-side signals, including lead times and order-arrival
co-movement.

\item We characterize forecastability uniformly across KPI families, processes, and customers, contrasting classical baselines with Chronos and finding a recurring family-level ordering across both processes.
\end{itemize}

In the remainder, \autoref{sec:canonical} frames the event log as a canonical
form over the relational ERP schema; \autoref{sec:heterogeneity} uses that
representation to characterize heterogeneity in the customer base;
\autoref{sec:extraction} defines the three KPI families as uniform operations
over it; and \autoref{sec:crossprocess} and \autoref{sec:forecastability}
study cross-process comparability and forecastability of the resulting KPI time
series. \autoref{sec:limitations} discusses limitations. \autoref{sec:conclusion} concludes.

\section{The Event Log as a Canonical Form}
\label{sec:canonical}

An event log can be viewed as a nested table whose rows are cases (see
\autoref{fig:canonical}). Each case identifies a leading business document
(e.g., a sales order) and carries case-level attributes such as customer and net
value; associated with each case is an ordered sequence of event records. Each
event has an activity label from a controlled vocabulary, a timestamp, and
optional payload such as resource. Here, an \emph{event} is a recorded occurrence in the log, while its
\emph{activity} is the label of the processing step that occurrence represents. Standards such as XES~\cite{xes2016}
standardize how such logs are structured and exchanged, but the format alone
does not guarantee that two customers use the same label for the same processing
step.

\begin{figure*}[!htb]
\centering
\includegraphics[width=\textwidth]{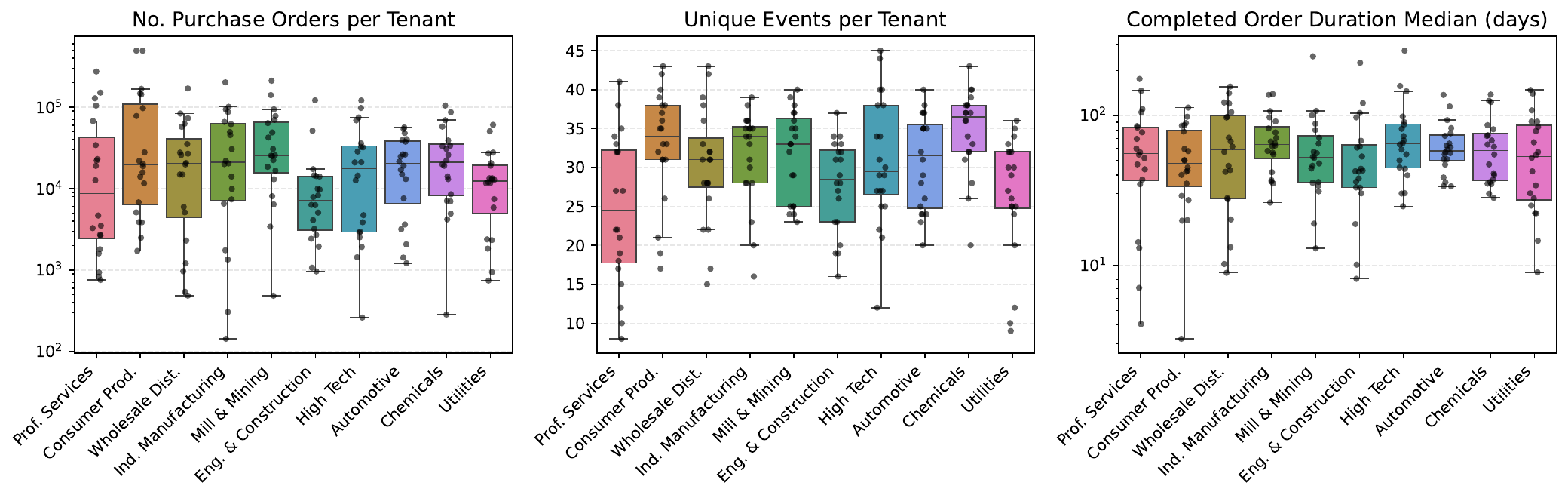}
\caption{Volume, complexity, and duration per customer, by industry. Twenty customers are shown per industry. The same
process varies by orders of magnitude across customers.}
\label{fig:heterogeneity}
\end{figure*}

Two things make the labels comparable across customers. First, all customers in our work run S/4HANA Cloud\footnote{Each customer is a
separate S/4HANA Cloud tenant; we use \emph{customer} and \emph{tenant}
interchangeably throughout, including in the figures.}, so the underlying
tables, fields, and coded values are largely standardized by default. Second, a fixed mapping---the shared
activity ontology---turns that shared substrate into a common vocabulary: each activity is defined by an extraction rule that selects the relevant source
records and maps their event-defining field or code value to a canonical label.
Different steps draw on different sources---a goods movement, for instance, is
labeled by its movement type (a goods receipt becomes \emph{Create Goods
Receipt}), while a field change recorded in a change document (CDPOS) is labeled
by the changed field (a change to the net-price field becomes \emph{Change PO
Item Net Price}). This comparability holds only where tenants use the standard
fields and codes assumed by the rules; customized fields or codes may therefore
be missed or mislabeled rather than harmonized.

Each rule also carries a timestamp: the appropriate
native timestamp for
transactional documents and the change-log time for a field
change. The mapping therefore assigns a single event time to
each activity occurrence, regardless of whether the underlying
evidence comes from a transactional document or a change
document. Applying the same rule library uniformly across tenants therefore aligns not
only the labels but the event times, so the resulting series are comparable
across customers.

Beyond semantic comparability, the event log also has a flat
relational view: tuples over (case, activity, timestamp) and
payload attributes, on which relational algebra applies directly.
It is a denormalized reshape of the normalized ERP source,
gathering facts scattered across order, delivery, invoice, and
payment tables under one schema. The ontology makes activity
labels comparable; the common schema makes KPI operators
reusable; and centralizing source-specific joins in the one-time
mapping yields extraction leverage.


\section{The Customer Base: Process Heterogeneity}
\label{sec:heterogeneity}

Before extracting KPI time series, we characterize the customer base through the
event-log representation itself. From (case, activity, timestamp) alone, three
per-customer descriptors follow directly: the number of cases (process volume),
the number of distinct activities (process complexity), and the median completed-case duration (order creation to final payment).
Fig.~\ref{fig:heterogeneity} reports these descriptors grouped by industry. The
same end-to-end process is executed very differently across customers: volumes
span two to three orders of magnitude within a single industry, while
complexity and typical durations vary systematically across industries.

\begin{figure*}[t]
\centering
\includegraphics[width=0.9\textwidth]{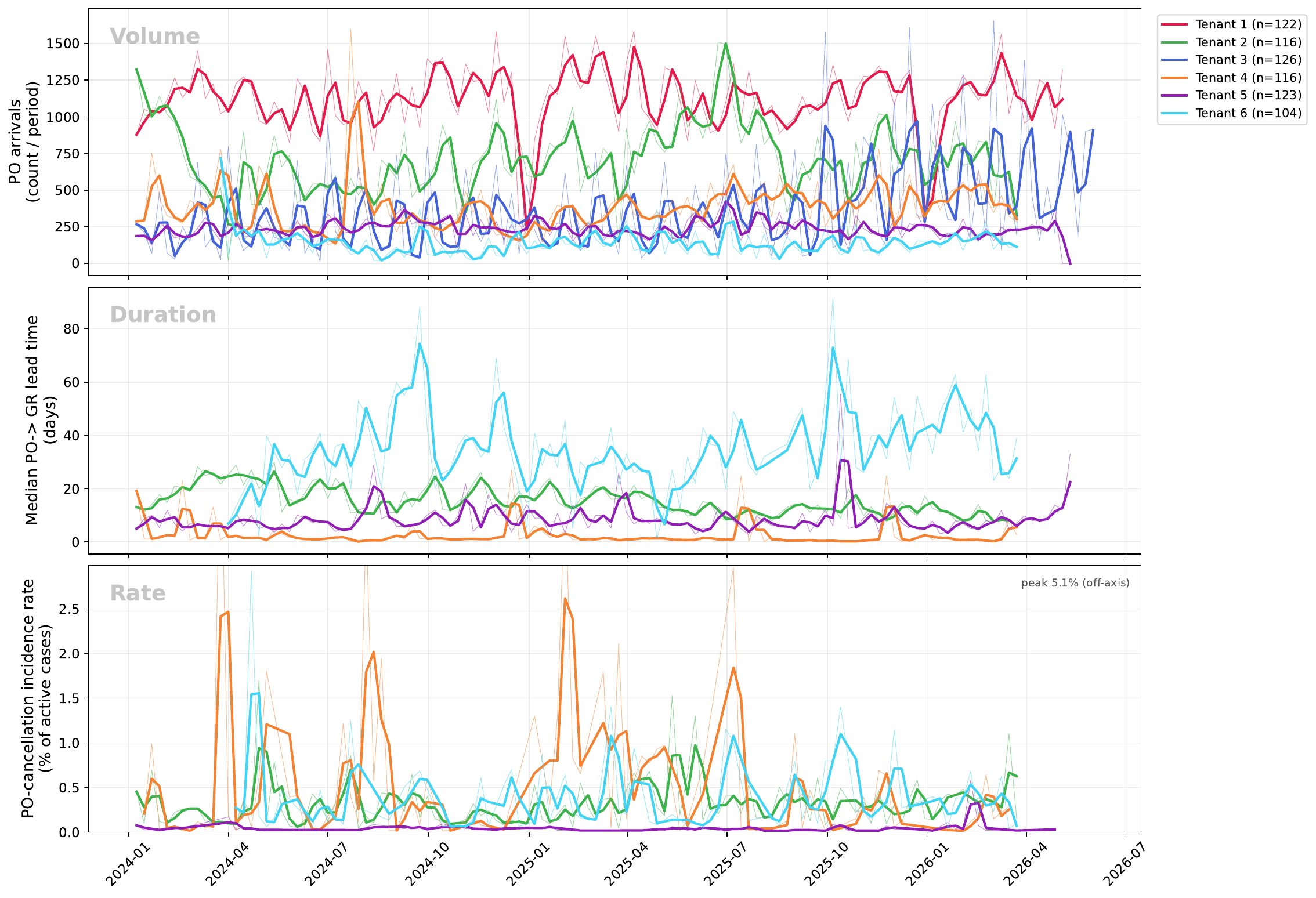}
\caption{The three KPI families extracted per customer for
procure-to-pay: volumes (arrivals), durations (purchase-order
to goods-receipt lead time), and active-case-normalized rates
(cancellation rate). For readability, the figure shows the six
tenants with the largest ordered quantities in one industry.
Bold lines are smoothed via a 2-point rolling mean.}
\label{fig:families_ts}
\end{figure*}

\begin{figure*}[t]
\centering
\includegraphics[width=0.9\textwidth]{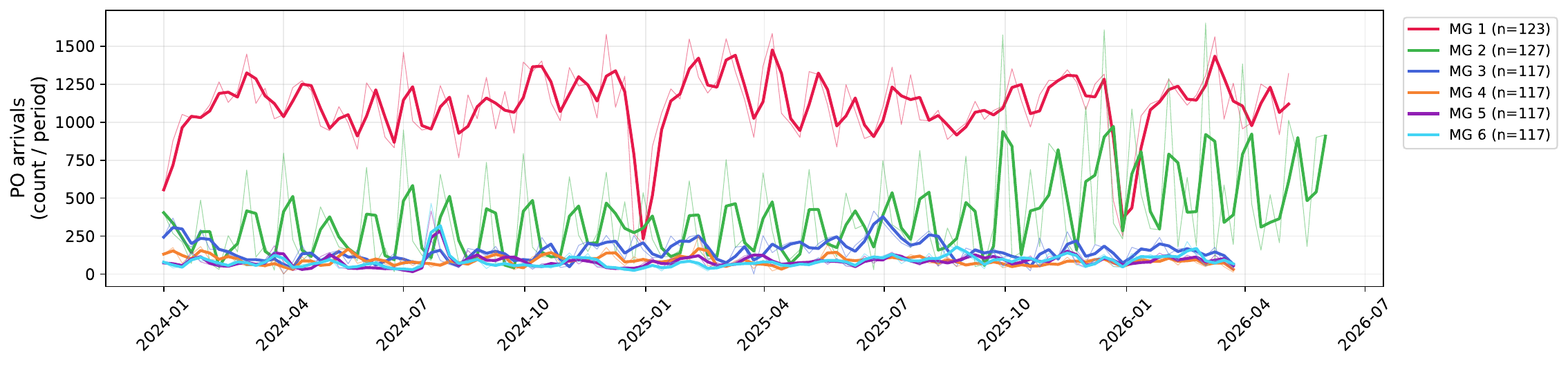}
\caption{Order arrivals decomposed to material-group level for six material groups across three tenants (each tenant has multiple material groups), illustrating how the event-log representation supports drill-down without an additional join.}
\label{fig:material}
\end{figure*}

\section{KPI Extraction}
\label{sec:extraction}
We group process-derived KPIs into three families, each defined by a single operation over
the event log (Table~\ref{tab:families}). \emph{Volumes} count cases for an activity
per period (order arrivals, completions). \emph{Durations} take the timestamp
difference between two activities within a case (e.g., purchase-order to goods
receipt, sales-order to delivery); these are process lead times.
\emph{Rates} count distinct cases with an activity per period relative to an active-case population (e.g., cancellations, price changes).

\begin{table}[htbp]
\caption{KPI families as operations over the canonical form}
\label{tab:families}
\centering
\renewcommand{\arraystretch}{1.3}
\begin{tabular}{p{1.1cm}p{2.7cm}p{2.7cm}}
\toprule
\textbf{Family} & \textbf{Event-log operation} & \textbf{Relational equivalent} \\
\midrule
Volume & Count of cases for an activity per period & Aggregate the activity's timestamp; e.g., native for arrivals, resolved from the terminal event for completions. \\
Duration & Timestamp difference between two activities per case, bucketed by end period & Join the start and end timestamp sources through the
schema-specific key path that identifies a common case; a
separate join path per metric \\
Rate & Distinct cases with the activity per period over the active-case denominator & Gather qualifying events; count distinct cases per period over a temporal open-case denominator\\
\bottomrule
\end{tabular}
\end{table}

Volume metrics are bucketed at case arrival or completion. Rate metrics are bucketed at the event that defines them---for
example, a price change or a rejection---and attributed to the
period in which that event was recorded. Duration metrics are
bucketed at the terminal event of each interval rather than at its start, so
every period contains only cases with a complete, observed duration and the
series is not right-censored at the cost of a completion-time lag.

As raw counts, rate numerators differ considerably across customers: a large customer cancels more orders simply because it processes more. The numerator counts distinct cases with a qualifying event, so a case contributes at most once per period. We normalize it by active cases—cases from arrival until completion, or until 12 weeks after their last event if no completion is observed. The result is an active-case-normalized event rate, not a probability. This denominator is itself a byproduct of the event-log representation:
openness is a predicate over a case’s observed activity sequence and can therefore be evaluated uniformly across processes. Over the raw schema, the same quantity would require repeatedly reconstructing, per case and per period, whether a start event had occurred without a corresponding completion event across multiple timestamp sources.

All metrics are computed at weekly granularity and all series are screened before analysis. 
We require at least 30 cases per period (the denominator for rates, the sample size for a duration median) and a minimum of 100 weekly periods per series (criteria and counts in Table~\ref{tab:dataset}); series below these thresholds are held out and reported as coverage caveats rather than analyzed. Reported counts (n) are post-screening. Fig.~\ref{fig:families_ts} shows one representative KPI from each family extracted per customer for procure-to-pay; the same extraction, unchanged, runs on order-to-cash.

Recovering these metrics over the raw schema is bespoke per family: A single duration, for instance, may need to recover its start and end timestamps
from separate sources and match them on a case identity that
differs across schemas; a different duration then requires a
different join path. Over the event-log, by contrast, every
volume is the same count, every duration the same within-case
timestamp difference, and every rate the same normalized
count. Table~\ref{tab:leverage} traces two examples.

Because the common structure retains each order's material reference, any KPI
can be decomposed to the material-group level without a new join.
Fig.~\ref{fig:material} shows order arrivals decomposed to the material-group level for the top materials by ordered quantity in three customers; the resulting per-material series act as demand proxies and expose structure that the customer-level aggregate dampens.

Some representative activity correspondences used to create these KPIs in order-to-cash and procure-to-pay are listed in
Appendix~\ref{app:activities}.


\begin{figure*}[t]
\centering
\begin{subfigure}[b]{0.36\textwidth}
  \centering
  \includegraphics[width=\linewidth]{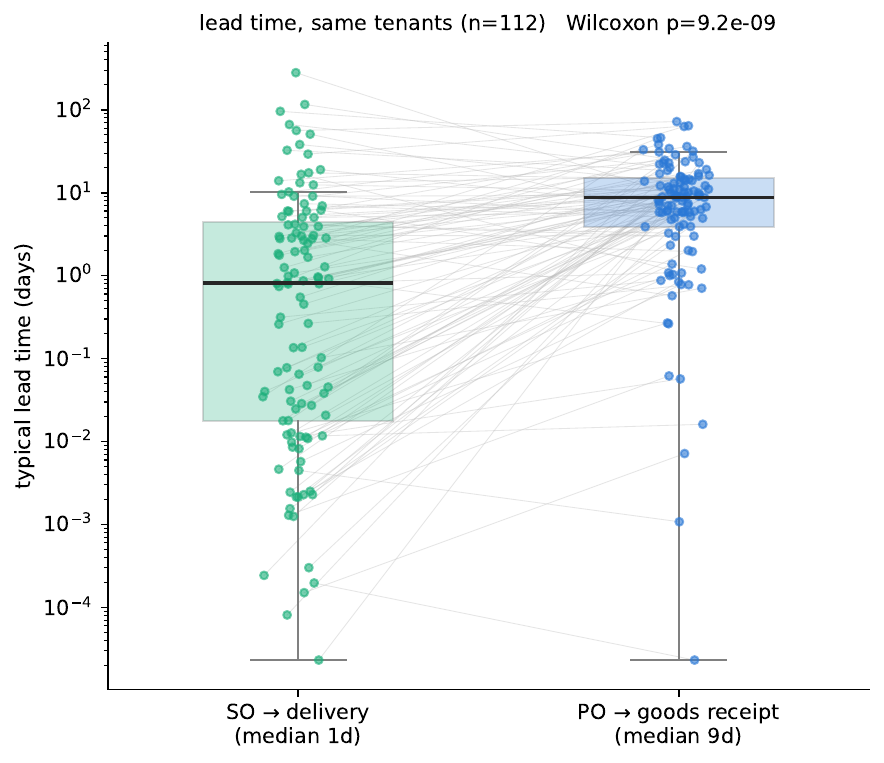}
  \caption{}
  \label{fig:leadtime}
\end{subfigure}\hfill
\begin{subfigure}[b]{0.62\textwidth}
  \centering
  \includegraphics[width=\linewidth]{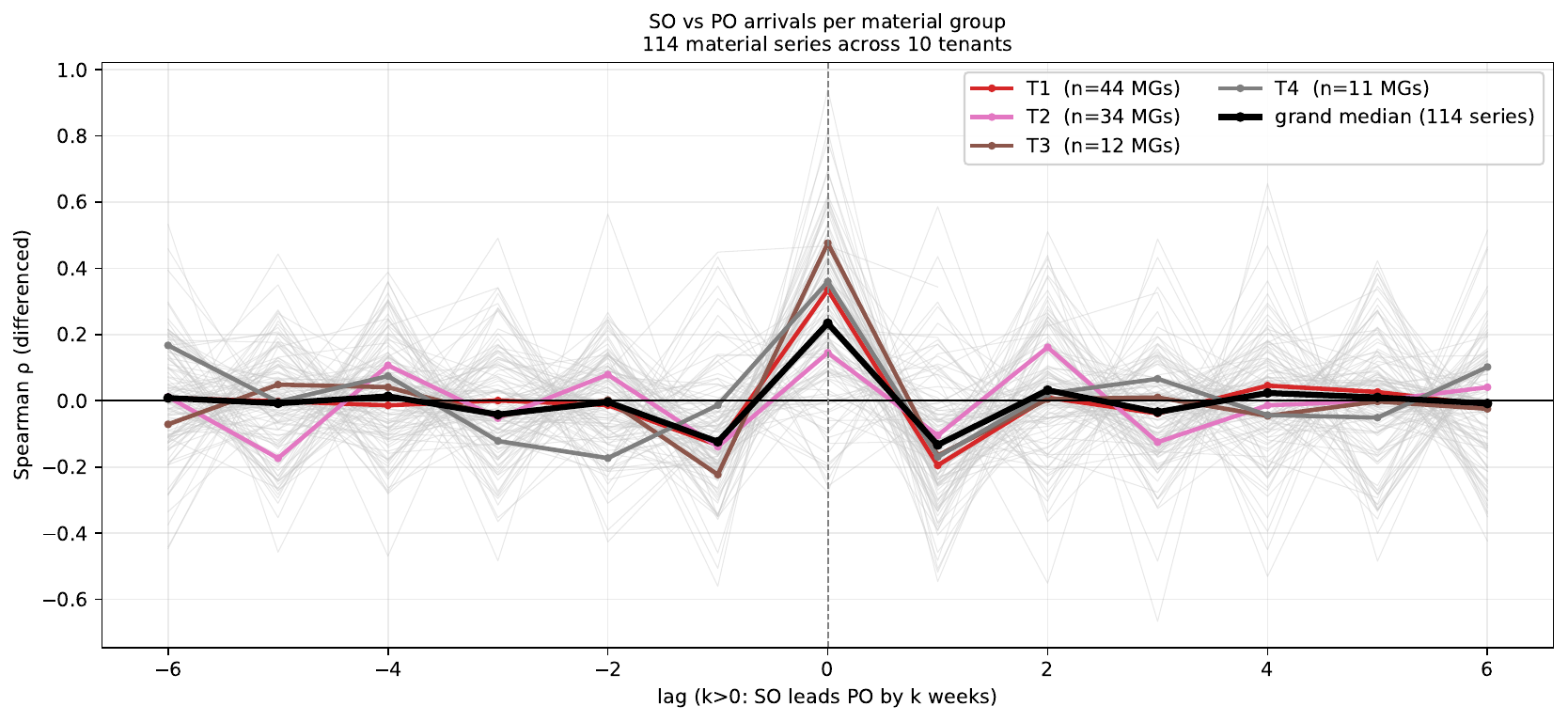}
  \caption{}
  \label{fig:xcorr}
\end{subfigure}
\caption{Cross-process comparison of sales (SO) and procurement (PO) sides.
(a) Paired customer-level median lead times for customers where both are
observed; procurement is longer than sales (median 9 vs.\ 1 day, 
Wilcoxon $p<10^{-8}$). (b) Lagged cross-correlation of weekly SO vs.\ PO
arrivals per material group for the wholesale distribution industry; the peak at
lag zero indicates contemporaneous co-movement with no systematic lead or lag.
Four tenant medians are highlighted.}
\label{fig:crossprocess}
\end{figure*}
\section{Cross-Process Comparison}
\label{sec:crossprocess}

Because order-to-cash and procure-to-pay reduce to the same common structure,
their signals become directly comparable, a comparison that over raw ERP would
require reconciling two unrelated join paths. We use this comparability in two
ways: first, to compare customer-level lead-time summaries across the two
processes; and second, to compare the weekly arrival series themselves.

Fig.~\ref{fig:leadtime} compares customer-level median lead times on the sales
side (sales order to delivery) and the procurement side (purchase order to goods
receipt) for customers where both are observed. Procurement lead times are
markedly longer (median 9 vs.\ 1 day; n=112 tenants, Wilcoxon
test, $p<10^{-8}$), consistent with supplier lead times reflecting more external dependencies than internal fulfillment.

Fig.~\ref{fig:xcorr} then turns to the time-series view, showing lagged cross-correlation
between weekly sales- and procurement-side arrivals by material group in
wholesale distribution; the material-group match is a coarse proxy for shared
demand exposure, as a group bundles items that may be sold and procured as
distinct materials. The cross-correlation peaks moderately at lag zero
(grand-median $\rho\approx0.24$) with no systematic lead or lag in either
direction. Individual customers vary in the strength of this coupling, and a
few show weak off-zero structure, but no consistent lead or lag replicates
across the customer base: the two sides appear synchronized rather than
sequential at weekly resolution, and finer coupling is not detectable here.
This co-movement is consistent with, but does not by itself establish, a
replenishment link, since common drivers such as calendar or market conditions
produce the same pattern.

\section{Forecastability}
\label{sec:forecastability}

The same shared representation also makes forecastability comparable across
KPI families, processes, and customers, letting us ask which ERP KPIs carry
predictive structure beyond persistence.

\begin{figure*}[t]
\centering
\includegraphics[width=\textwidth]{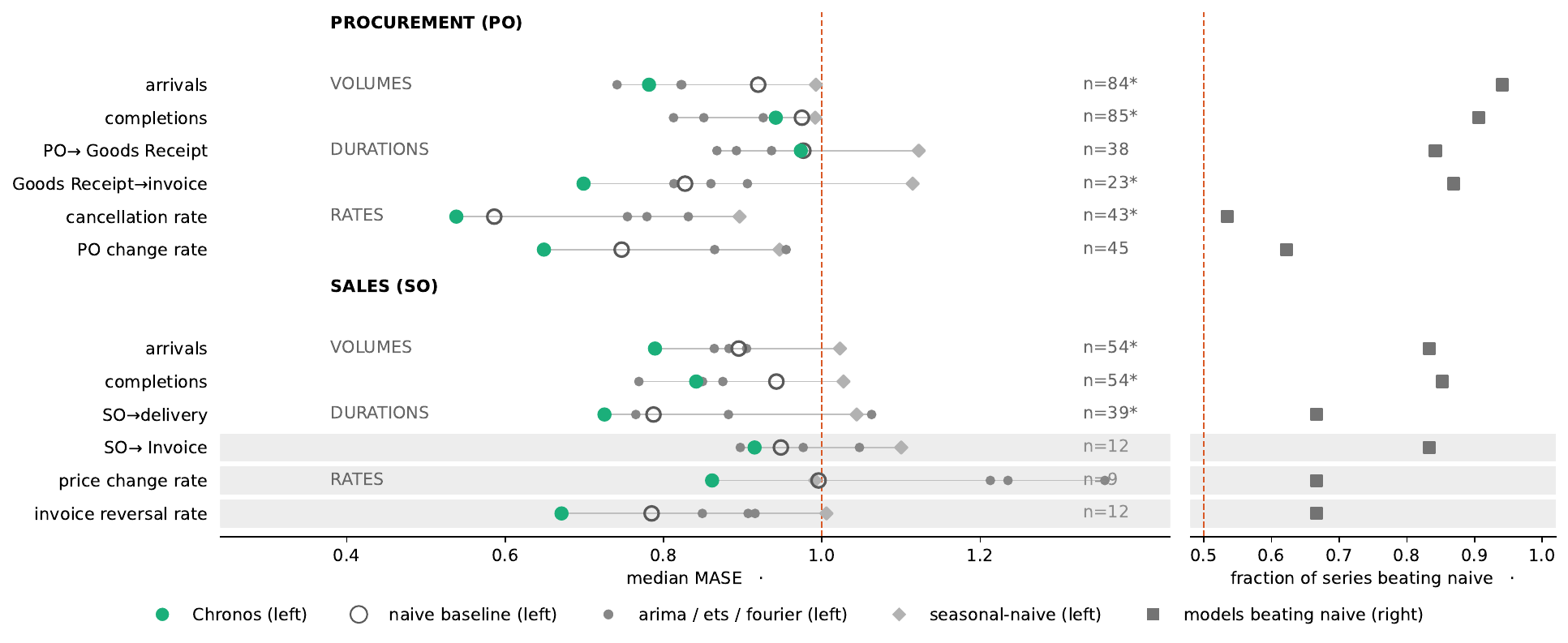}
\caption{Forecastability by process and family. Left: median MASE per model
(lower is better). Right: fraction of series where any model beats the na\"{\i}ve
baseline. Thin rows (small sample) are shaded. $^*$ means at least one model beats na\"{\i}ve on the same series (one-sided Wilcoxon
   signed-rank, Holm-corrected over all models and KPIs within process and
   family, $p<0.05$).}
\label{fig:forecastability}
\end{figure*}

We assess forecastability with a rolling-origin
evaluation~\cite{tashman2000outofsample} and mean absolute scaled error
(MASE)~\cite{hyndman2006another}, i.e., error relative to an in-sample one-step
na\"{\i}ve forecast. Gaps of at most four weeks are linearly interpolated;
series with longer gaps are excluded. Cut-offs are four weeks apart; models
train on all prior weeks (at least 80) and forecast the next 12, giving at
least three evaluations per series, scored by their median MASE. Models are
ETS with additive trend~\cite{hyndman2002state}; ARIMA~\cite{box2015time}
($p,q\leq3$, $d\in\{0,1\}$ by an ADF test, order by AIC on the first training
window); the same ARIMA order with $K=3$ Fourier terms at
$m=52$~\cite{pankratz2012forecasting}; seasonal na\"{\i}ve ($m=52$); and
Chronos (\texttt{chronos-t5-small})~\cite{chronos2024}, applied zero-shot with
the per-horizon median of 20 samples. Classical models are refit at each
cut-off; no selection or tuning uses test windows.
Fig.~\ref{fig:forecastability} reports median MASE per KPI (left) and the
fraction of series on which any model beats the out-of-sample na\"{\i}ve
baseline (right), grouped by family, separately for procurement and sales.

Three patterns hold across both processes. First, seasonal-na\"{\i}ve is
the weakest model almost everywhere (rightmost on the left panel), suggesting
that simple annual repetition is not the dominant source of predictability at
weekly resolution. Second, volume series are the most forecastable family: some
model beats the na\"{\i}ve baseline on roughly 83--94\% of series, and both
classical models and Chronos often improve on the na\"{\i}ve baseline, so
arrivals and completions carry clear structure beyond persistence. Third,
forecastability decreases from volumes through durations to rates; the
beat-na\"{\i}ve fraction for several rate families is only marginally above one
half, so for rates the na\"{\i}ve baseline is substantially harder to improve
on. Paired tests support this ordering: after Holm correction over all models and
KPIs within each process and family, at least one model improves significantly
on na\"{\i}ve for all four volume KPIs, two of four durations, and one of four rates (Fig.~\ref{fig:forecastability}\textsuperscript{*}).

This ordering recurs across the independently extracted procurement and
sales processes, and holds across the highly non-uniform customer population of
Fig.~\ref{fig:heterogeneity}, suggesting that forecastability aligns more
strongly with KPI family than with process alone.

Across models, the comparison is family-dependent. On volumes, the classical
models hold a small but consistent edge over Chronos in median MASE on both
processes. On durations the two are close, with Chronos at or near the best
model on most families, though not on purchase-order to goods receipt.

Rates\footnote{v2 (arXiv): the rate labeled \emph{receipt rate} in the previous version of Fig.~\ref{fig:forecastability} is in fact the PO \emph{change rate}. We correct the label and clarify the rate definition accordingly. Results are unchanged.}, at the bottom of that ordering, were the family we expected to be
least forecastable: cancellations, price changes, and reversals are event-like
shocks with little obvious persistence, so a priori the na\"{\i}ve baseline
should be difficult to beat. The results largely bear this out. Yet the margin
above na\"{\i}ve is not zero, and on the rate families that are forecastable at
all, Chronos is often the strongest model (leftmost on cancellation and
price-change rate). This observation rests on thin families---two of the four
rate families have $n \leq 12$ (price-change rate $n=9$, invoice-reversal rate
$n=12$)---so we report it as suggestive rather than conclusive.

We use active-case counts only as normalization denominators and do not assess them as forecasting targets.

\section{Discussion and Limitations}
\label{sec:limitations}
This paper does not come with a released benchmark as we legally cannot publish customer
transactions. However, the reported analyses use aggregated KPI series, and
the results are reproducible within our environment. Rate families are thin for some sales-side metrics
and we tier claims by sample size accordingly. 
The material-group match used for cross-process comparison
is a coarse proxy for shared demand exposure, not an item-level
inventory or replenishment link. A supply-chain analysis of
demand propagation or the bullwhip effect\cite{lee1997bullwhip}  would require
joining sales demand, replenishment orders, inventory positions,
and supply relationships at item level; these joins are often
sparse and are left to future work. Table \ref{tab:dataset} documents dataset and post-screening coverage, and Table \ref{tab:leverage} makes the schema-specific extraction burden concrete for two representative KPIs. We do not quantify per-tenant ontology mapping coverage or perform a sensitivity analysis of the screening thresholds; both are left to an extended version.


\section{Conclusion}
\label{sec:conclusion}
In this paper, we framed the event log as a common representation for KPI extraction over relational ERP systems. Its value is extraction leverage: once mapped, process-derived KPIs reduce to reusable operations over one structure rather than bespoke joins over many tables. Using this representation, we defined three KPI families and characterized their heterogeneity, cross-process comparability, and forecastability across hundreds of customers.
\FloatBarrier                 
\bibliographystyle{IEEEtran}
\bibliography{references}
\appendices

\section{Representative Activities for KPI Construction}
\label{app:activities}
Table~\ref{tab:activity_mapping} lists some of the representative activities in the
shared activity ontology for order-to-cash (O2C) and procure-to-pay (P2P).
Durations are shown as \emph{start}~$\rightarrow$~\emph{end} activity pairs.
The listed activities are illustrative and not exhaustive.

\section{Dataset Statistics}
\label{app:dataset}
Table~\ref{tab:dataset} summarizes the event logs and the KPI series derived
from them. Log statistics cover all tenants with an extracted log; KPIs are
computed on the tenant subsets listed under \emph{KPI series} and screened
per (tenant, KPI) series as described in the table notes.

\section{Extraction effort}
\label{app:extraction}
Table~\ref{tab:leverage} contrasts two KPIs from Fig.~\ref{fig:forecastability}.
Over the raw schema, each requires its own multi-table join path, and the rate requires a separate denominator reconstruction; over the event log, both are expressed by generic operator templates instantiated with the relevant activities.

\begin{table*}[t]
\caption{Examples of representative activities used for KPI construction in O2C and P2P.
Durations are given as start~$\rightarrow$~end pairs. Some metrics aggregate multiple related activities (e.g.,
the change/rework rate spans the price, quantity, and material variants).}
\label{tab:activity_mapping}
\footnotesize
\setlength{\tabcolsep}{3pt}
\renewcommand{\arraystretch}{1.0}
\begin{tabular}{p{0.13\textwidth} p{0.14\textwidth} p{0.33\textwidth} p{0.33\textwidth}}
\toprule
\textbf{Family} & \textbf{KPI} & \textbf{O2C activities} & \textbf{P2P activities} \\
\midrule
\multirow{2}{*}{Volume}
& case start   & Create Sales Order Item & Create PO  \\
& case completion   & Clear Invoice & Clear Invoice \\
\midrule
\multirow{3}{*}{Duration}
& lead time
& Create Sales Order Item $\rightarrow$ Create Delivery
& Create PO Item $\rightarrow$ Create Goods Receipt \\
& goods\,$\rightarrow$\,invoice
& Post Goods Issue $\rightarrow$ Create Invoice
& Create Goods Receipt $\rightarrow$ Record Invoice Receipt \\
& lifecycle
& Create Sales Order Item $\rightarrow$ Clear Invoice
& Create PO Item $\rightarrow$ Clear Invoice \\
\midrule
\multirow{4}{*}{Rate}
& change/rework
& Change Sales Order Item Price
& Change PO Item Net Price \\
& reschedule
& Change Sales Order Reqd Deliv Date
& Change PO Requested Delivery Date \\
& cancellation
& Reject Sales Order Item
& Delete PO Item \\
& reversal/return
& Create Invoice Cancellation; Create Return Delivery
& Reverse Goods Receipt; Return Delivery  \\
\bottomrule
\end{tabular}
\end{table*}

\begin{table}[t]
\centering
\caption{Dataset statistics; median [IQR] across tenants unless noted.}
\label{tab:dataset}
\small
\setlength{\tabcolsep}{4pt}
\begin{tabular}{@{}lrr@{}}
\toprule
 & O2C & P2P \\
\midrule
\multicolumn{3}{@{}l}{\emph{Event logs}} \\
Tenants / industries           & 694 / 19 & 748 / 19 \\
Cases per tenant               & 8.3k [1.8k, 47.7k] & 9.1k [2.5k, 26.3k] \\
Events per tenant              & 63k [12k, 341k] & 61k [15k, 165k] \\
Distinct activity labels observed             & 69 & 49 \\
Observation window             & 2017--2026 & 2017--2026 \\
Incomplete cases (\%)$^a$          & 46 [25, 75] & 52 [33, 82] \\
\midrule
\multicolumn{3}{@{}l}{\emph{KPI series}} \\
Tenants / industries$^b$       & 220 / 11 & 311 / 16 \\
Tenants with kept series   & 57 & 86 \\
Series kept: vol.\ / dur.\ / rate & 108 / 51 / 21 & 169 / 61 / 88 \\
Kept length (weeks)$^c$        & 156 [100, 369] & 130 [100, 444] \\
\bottomrule
\end{tabular}

\vspace{2pt}
\parbox{\columnwidth}{\footnotesize
 Tenants with $<$500 events
are excluded. Periods with $<$30 cases are masked; a series is kept if it spans
$\geq$100 weeks, is $\geq$80\% observed, has no gap $>$4 weeks, and takes
$\geq$5 distinct values. Per-KPI counts are shown in Fig.~\ref{fig:forecastability}.
$^a$ “Completion” refers to the terminal event used for the corresponding process: for O2C: \textit{Clear Invoice} or \textit{Sales Order Completely Processed}, for P2P: \textit{Clear Invoice}
$^b$\,O2C: 20 tenants per industry for industries with $\geq$20 tenants;
P2P: all tenants available at extraction time. Numbers are lower than total event logs, due to removal of low volume tenants.
$^c$\,Median [min, max].}
\end{table}

\begin{table}[t]
\centering
\caption{Illustrative raw-schema extraction paths versus canonical event-log operations for two KPI definitions}
\label{tab:leverage}
\footnotesize
\setlength{\tabcolsep}{3pt}
\begin{tabular}{@{}>{\raggedright\arraybackslash}p{0.55in}
                  >{\raggedright\arraybackslash}p{1.95in}
                  >{\raggedright\arraybackslash}p{0.9in}@{}}
\toprule
KPI & Raw schema & Event log \\
\midrule
SO$\to$invoice  (O2C)
& VBAP $\to$ LIPS $\to$ VBRP via two alternative paths (order- or
  delivery-related billing) $\to$ BKPF/BSEG via a six-field key fallback
  $\to$ clearing document via \texttt{AUGBL}; at least 8 tables, $\geq$11 joins
& $t(\textit{Clear Invoice})-t(\textit{Create SO Item})$ \\
\addlinespace
Cancellation rate (P2P)
& Numerator: CDHDR$\bowtie$CDPOS rows on EKPO with field \texttt{LOEKZ}
  set to L/X (deletion); PO item parsed from \texttt{TABKEY}
  (client\,+\,PO\,+\,item). Denominator (open POs): from creation
  (CDPOS insert record; EKPO stores only a date) until clearing of the
  vendor invoice (EKBE $\to$ BSEG $\to$ \texttt{AUGBL});
  7 tables, $\geq$8 joins
& cases with \textit{Delete PO Item} $/$ active cases \\
\bottomrule
\end{tabular}
\end{table}


\end{document}